\documentclass[aps, prb, twocolumn, 10pt, superscriptaddress]{revtex4-2}

\usepackage{graphicx}
\usepackage{siunitx}
\usepackage{upgreek}
\usepackage{amssymb,amsmath}
\usepackage{xcolor}

\usepackage{amssymb}
\usepackage{gensymb}
\usepackage{graphicx}
\usepackage{bm}
\usepackage{tikz}
\usepackage{siunitx}
\usepackage{amsmath}

\def\be{\begin{equation}}
\def\ee{\end{equation}}
\def\bea{\begin{eqnarray}}
\def\eea{\end{eqnarray}}
\def\e{\epsilon}

\def\t{\tilde}

\begin{document}

\title{An intermediate morphology in the patterning of the crystalline Ge(001) surface induced by ion irradiation}

\author{Denise J. Erb}
\email[]{d.erb@hzdr.de}
\affiliation{Ion Beam Center, Institute of Ion Beam Physics and Materials Research, Helmholtz-Zentrum Dresden-Rossendorf, Dresden, Germany}

\author{Daniel A. Pearson}
\affiliation{Division of Science and Engineering, Pennsylvania State University, Abington, Abington, PA, 19001, USA}

\author{Tom\'{a}\v{s} \v{S}kere\v{n}}
\affiliation{IBM Research - Z\"{u}rich, S\"{a}umerstrasse 4, 8803 R\"{u}schlikon, Switzerland}
\affiliation{Czech Technical University in Prague, Faculty of Nuclear Sciences and Physical
 Engineering, B\v{r}ehov\'{a} 7, 115 19 Prague 1, Czech Republic}

\author{Martin Engler}
\affiliation{Ion Beam Center, Institute of Ion Beam Physics and Materials Research, Helmholtz-Zentrum Dresden-Rossendorf, Dresden, Germany}
\affiliation{DRK Kliniken Berlin, Berlin, Germany}

\author{R. Mark Bradley}
\email{mark.bradley@colostate.edu}
\affiliation{Departments of Physics and Mathematics, Colorado State University, Fort
Collins, CO 80523, USA}

\author{Stefan Facsko}
\affiliation{Ion Beam Center, Institute of Ion Beam Physics and Materials Research, Helmholtz-Zentrum Dresden-Rossendorf, Dresden, Germany}
\homepage[]{www.hzdr.de/ibc}

\date{\today}

\begin{abstract}

We investigate the morphologies of the Ge(001) surface that are produced by bombardment with a normally incident, broad argon ion beam at sample temperatures above the recrystallization temperature.  Two previously-observed kinds of topographies are seen, i.e., patterns consisting of upright and inverted rectangular pyramids, as well as patterns composed of shallow, isotropic basins.  In addition, we observe the formation of an unexpected third type of pattern for intermediate values of the temperature, ion energy and ion flux. In this type of intermediate morphology, isolated peaks with rectangular cross sections stand above a landscape of  shallow, rounded basins.  We also extend past theoretical work to include a second order correction term that comes from the curvature dependence of the sputter yield.  For a range of parameter values, the resulting continuum model of the surface dynamics produces patterns that are remarkably similar to the intermediate morphologies we observe in our experiments. The formation of the isolated peaks is the result of a term that is not ordinarily included in the equation of motion, a second order correction to the curvature dependence of the sputter yield.

\end{abstract}

\maketitle

\section{Introduction
\label{Introduction}}

Low energy ion irradiation of solid surfaces often leads to the formation of self-organized patterns \cite{Navez1962, Facsko1999, Munoz-Garcia2014}. Since it was first discovered \cite{Navez1962}, this phenomenon has attracted a great deal of attention as a potential method of nanofabrication \cite{BuatierDeMongeot2009,Skeren2015}, but completely unraveling the complex underlying physics is still a matter of active research. It has become clear that there is not a single physical mechanism that governs the pattern formation on ion irradiated surfaces. Instead, a number of different erosive, redistributive, and diffusive effects act simultaneously, and which effects are dominant depends on the experimental conditions, i.e., on the surface temperature, ion incidence angle, ion mass and energy, etc.~\cite{Bradley1988, Valbusa2002, Hansen2006, Skeren2013}. 

In experiments in which an initially crystalline surface is bombarded with a broad noble gas ion beam, a layer at the surface of the material is amorphized by the impinging ions if the sample temperature $T$ is below the dynamic recrystallization temperature $T_R$. In contrast, if the target material is maintained at a temperature $T > T_R$, the damage done to the crystal structure by the ion impacts is rapidly annealed away, and the sample remains crystalline. To date, this effect of temperature has mainly been observed in semiconductors; the bulk diffusivity of metals is high enough that the defects produced by ion irradiation are efficiently annealed away even at low temperatures, and the crystal structure remains largely intact \cite{Costantini2001-1,Costantini2001-2,Valbusa2002,Michely2001,Kalff2001,Chason2010}. 

Ion erosion of a material that remains crystalline during ion bombardment at a given temperature has been argued to be analogous to molecular beam epitaxy (MBE) \cite{Costantini2001-1,Valbusa2002, Levandovsky2004, Ou2013}.  In the simplest version of this picture, the incident ions sputter away surface atoms and so produce vacancies in the crystal surface. These vacancies then diffuse on the crystal terraces until they attach to  step edges. The vacancies on the irradiated crystal surface are the analogs of the adatoms that diffuse over the growing surface of a crystal during MBE. Thus, erosion of the surface of an crystalline solid has been referred to as \lq\lq reverse epitaxy.'' In reality, this picture is too simple, since ion bombardment also displaces surface atoms without sputtering them away, producing adatoms in addition to surface vacancies. 

Both vacancies and adatoms diffusing on the surface are subject to the Ehrlich-Schwoebel (ES) barrier, which strongly influences the nanoscale pattern formation that can occur on the surface of crystalline materials during ion bombardment \cite{Golubovic2011,Valbusa2002,Levandovsky2004,Chason2010}. The ES barrier makes it more likely that an adatom (surface vacancy) attaches to an ascending (descending) step edge.  It therefore leads to an effective uphill mass current and tends to destabilize an initially flat surface. This current is typically anisotropic, which is a manifestation of the anisotropy of the underlying crystal lattice.  The ES barrier is also important during MBE, and in that case it is responsible for the mounding instability that sometimes occurs \cite{Golubovic2011,Levandovsky2004}.

Perhaps the most progress in understanding pattern formation on ion-eroded crystalline semiconductors has been made in the case of germanium targets. In 2013, Ou~{\it et al.}~bombarded a Ge(001) surface with a normally incident 1 keV argon ion beam at a selection of different sample temperatures $T$ \cite{Ou2013}.  For $T<T_R\cong 250^\circ$C, the sample surface was amorphized by the ion bombardment and remained flat.  A pattern of rectangular pyramids and inverted rectangular pyramids formed when the sample was maintained at a temperature not too far in excess of the recrystallization temperature, and the pattern coarsened in time.  Finally, for higher temperatures, a pattern of shallow basins separated by low ridges was observed.
Ou~{\it et al.}~also introduced a phenomenological theory that is able to account for the pattern formation they observed for $T>T_R$ \cite{Ou2013}. According to this theory, the role of the incident ions is simply to randomly produce vacancies and a smaller number of adatoms on the crystal surface and the instability is due to the ES barrier.

In this paper, we investigate the qualitatively distinct morphologies of the Ge(001) surface produced by normal-incidence ion irradiation at temperatures $T>T_R$, and explore the surface morphology's dependence on temperature, ion flux and ion energy.  We observe the two kinds of morphologies already seen by Ou {\it et al.}, i.e., patterns consisting of upright and inverted rectangular pyramids, as well as patterns composed of shallow, isotropic basins.  In addition, we observe the formation of an intermediate morphology for intermediate parameter values.  In this exotic type of pattern, structural elements of the anisotropic and isotropic patterns coexist:  
isolated peaks with sloped sides and rectangular cross sections stand above a landscape of shallow, rounded basins.  We also extend the existing continuum model of the surface dynamics to include a higher order effect of curvature dependent sputtering. For a range of parameter values, the resulting model produces surface patterns that are remarkably similar to the intermediate morphologies we observe in our experiments. The formation of the isolated peaks can be traced to a term in the equation of motion that would produce spike singularities if this were not prevented by the slope selection that comes from the ES effect.

It is important to note that patterns similar to our intermediate patterns were observed when a Pt(111) surface was bombarded with a normally incident 1 keV Xe ion beam \cite{Michely2001,Kalff2001}, although in the case of the Pt target, the peaks have sixfold rather than fourfold symmetry.  As far as we have been able to determine, however, this type of pattern has not previously been observed on a semiconducting target material.

This paper is organized as follows.  In Sec.~\ref{sct:Experiment}, we describe our experimental methods and then show the results of experiments carried out with a range of sample temperatures, ion energies and ion fluences.  We advance a model for the surface dynamics in Sec.~\ref{sec: theory}, carry out simulations of the pattern formation, and compare the results to the results of our experiments.  We discuss our findings and place them in context in Sec.~\ref{sec: discussion} and conclude in Sec.~\ref{sct:Conclusions}.

\section{Experiment
\label{sct:Experiment}}

\begin{figure*}
\includegraphics[]{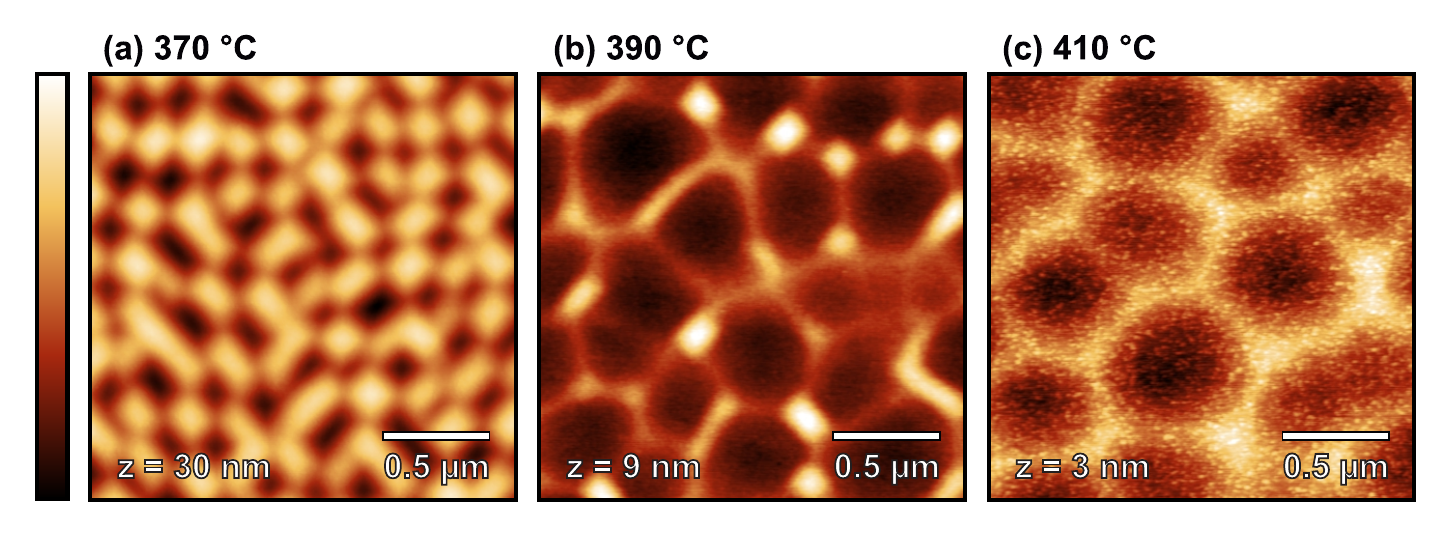}
\caption{AFM topographical images of Ge(001) surfaces after irradiation with 1000~eV $\text{Ar}^{+}$ ions with a flux of $f = 1 \times 10^{15}\text{cm}^{-2}\text{s}^{-1}$ at sample temperatures (a) \SI{370}{\degreeCelsius}, (b) \SI{390}{\degreeCelsius} and \SI{410}{\degreeCelsius}. The pattern morphology changes with increasing temperature, from an anisotropic pattern (a) to an intermediate morphology (b) and finally to an isotropic pattern (c). The false color ruler indicating the surface height ranges from 0 to $z$ as labeled.}
\label{fig:temperature-series}
\vspace{1cm}
\includegraphics[]{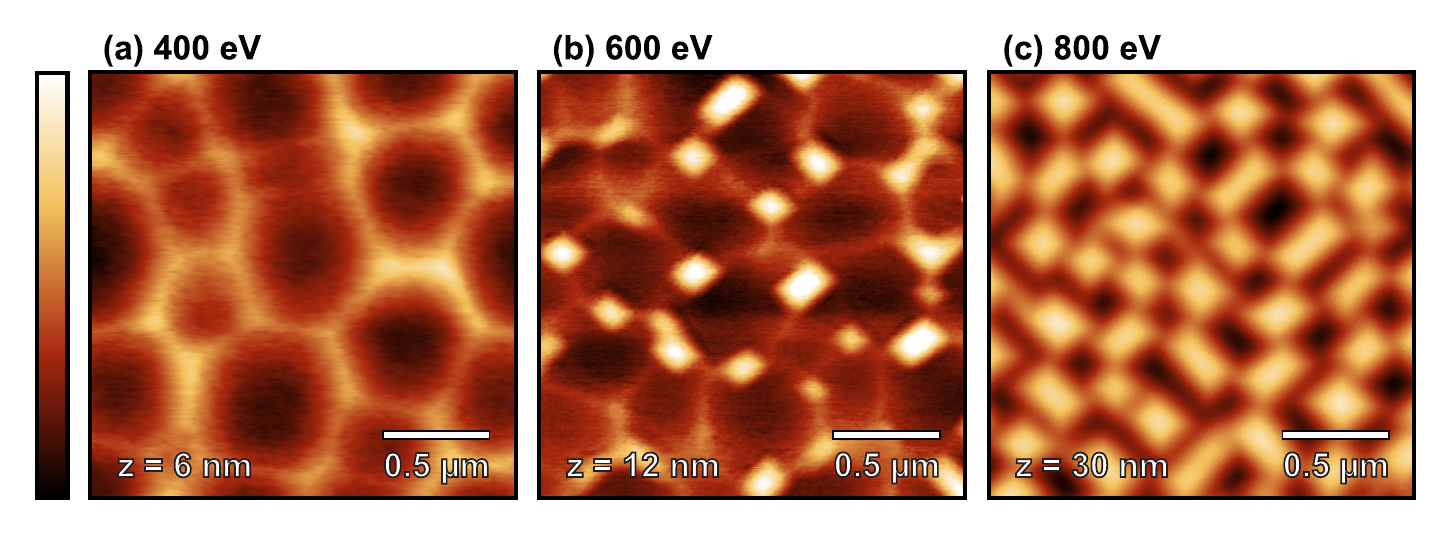}
\caption{AFM topographical images of Ge(001) surfaces after irradiation at a sample temperature of \SI{380}{\degreeCelsius} with $\text{Ar}^{+}$ ions with a flux of $f = 1 \times 10^{15}\text{cm}^{-2}\text{s}^{-1}$ and at ion energies (a) 400~eV, (b) 600 eV and 800~eV. The pattern morphology changes with increasing ion energy, from an isotropic pattern (a) to an intermediate morphology (b) and finally to an anisotropic pattern (c). The false color ruler indicating the surface height ranges from 0 to $z$ as labeled.}
\label{fig:energy-series}
\vspace{1cm}
\includegraphics[]{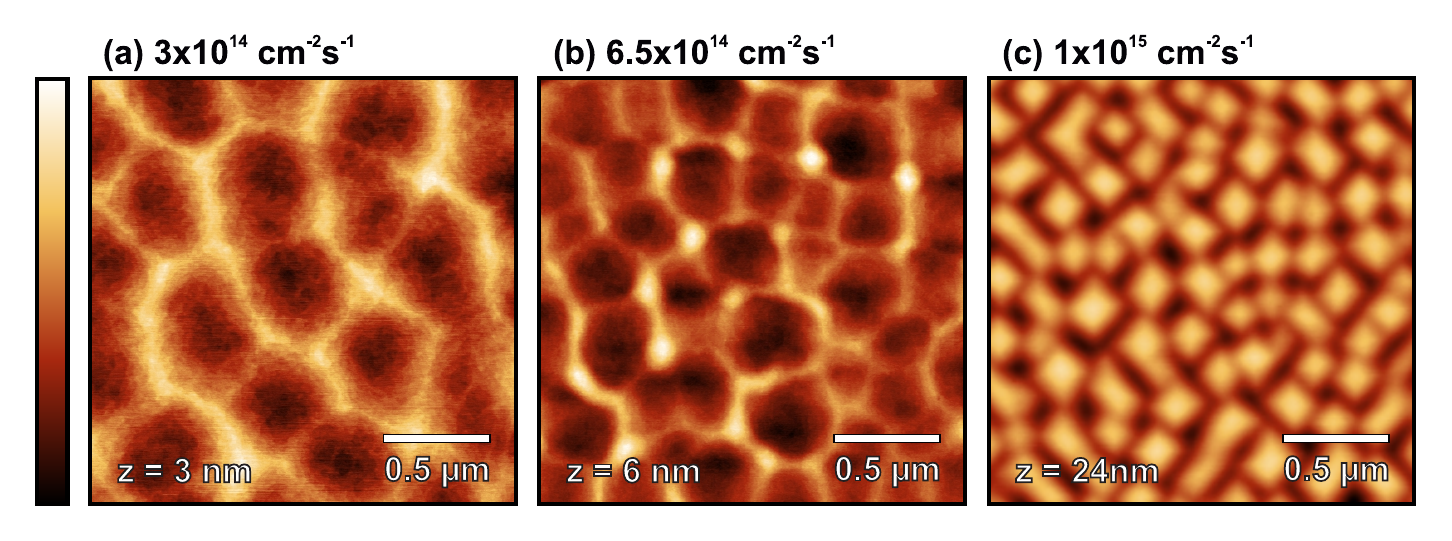}
\caption{AFM topographical images of Ge(001) surfaces after irradiation at a sample temperature of \SI{350}{\degreeCelsius} with $\text{Ar}^{+}$ ions with an energy of 400~eV at fluxes (a) $3 \times 10^{14}\text{cm}^{-2}\text{s}^{-1}$, (b) $6.5 \times 10^{14}\text{cm}^{-2}\text{s}^{-1}$ and (c) $1 \times 10^{15}\text{cm}^{-2}\text{s}^{-1}$. The pattern morphology changes with increasing ion flux, from an isotropic pattern (a) to an intermediate morphology (b) and finally to an anisotropic pattern (c). The false color ruler indicating the surface height ranges from 0 to $z$ as labeled.}
\label{fig:flux-series}
\end{figure*}

\subsection{Experimental Procedures and Data Processing
\label{ssct: Experimental Procedures and Data Processing}}

The samples were prepared by cleaving a commercial 2-inch Ge(001) wafer of 0.5~mm thickness into 10~mm $\times$ 10~mm pieces. The pieces were then mounted on a stainless steel plate by means of two narrow Ta strips, which held two opposite corners of each sample and were spot welded onto the plate. Each sample plate was transferred into a UHV system with base pressure $p_{0} < 1 \times 10^{-7}$~mbar and was clamped onto a stainless steel block mounted on a heated ceramic plate. The sample temperature was measured with a pyrometer pointing at the sample surface through a viewport at an angle of 45$^{\circ}$. Continuous temperature measurement was not possible due to the intense infrared radiation generated by the ion source during operation. Therefore, the temperature was stabilized before irradiation and then periodically checked for stability, and minor adjustments were made to the power supplied to the heater if necessary. Samples were irradiated with a broad beam of $\text{Ar}^{+}$ ions from a Kaufman type ion source at normal incidence with a working gas pressure of $p_{\text{Ar}} = 1.7 \times 10^{-4}$~mbar. The ion energy $E$ was varied in the range from 200~eV to 1000~eV, and the sample temperature was varied in the range from \SI{265}{\degreeCelsius} to \SI{410}{\degreeCelsius}.  The ion fluence was chosen so that the total amount of sputtered material was the same for all samples.  (The
energy-dependent sputter yields estimated by \textsc{SRIM} \cite{Ziegler1985}
were employed in making these choices.) This resulted in fluences between $F = 2.0 \times 10^{18}~\text{cm}^{-2}$ for 1000~eV and $5.4 \times 10^{18}~\text{cm}^{-2}$ for 200~eV $\text{Ar}^{+}$ ion irradiation. Atomic force microscope (AFM) step height measurements confirmed that the erosion depth was $(1400 \pm 105)$~nm for all samples. Sample characterization was performed ex-situ with a Bruker Multimode8 AFM in tapping mode. The topography data were processed, evaluated, and plotted using the Gwyddion software package \cite{Necas2012}. The characteristic lateral length scale $L$ was defined to be $2\pi / k_{\text{max}}$, where $k_{\text{max}}$ is the position of the maximum in the radial power spectral density.

Mean curvature images of both experimental and simulated data (see section \ref{sec: theory}) were created using Python and Matplotlib. The mean curvature $H$ of the surface $u=u(x,y)$ was calculated by the formula
\[ H(x,y)=\dfrac{(1+u_x^2)u_{yy}-2u_x u_y u_{xy}+(1+u_y^2)u_{xx}}{2(1+u_x^2+u_y)^{3/2}}\]
where the partial derivatives were evaluated using finite differencing. The units of $x$, $y$ and $u$ used for this calculation with experimental data were all microns, and so the units of $H$ were inverse microns. Prior to calculating the mean curvature for the experimental data, a small Gaussian smoothing filter (\verb|scipy.ndimage.gaussian_filter|) was applied to the surface $u$ in order to reduce noise.

\begin{figure*}
\includegraphics[width=\textwidth]{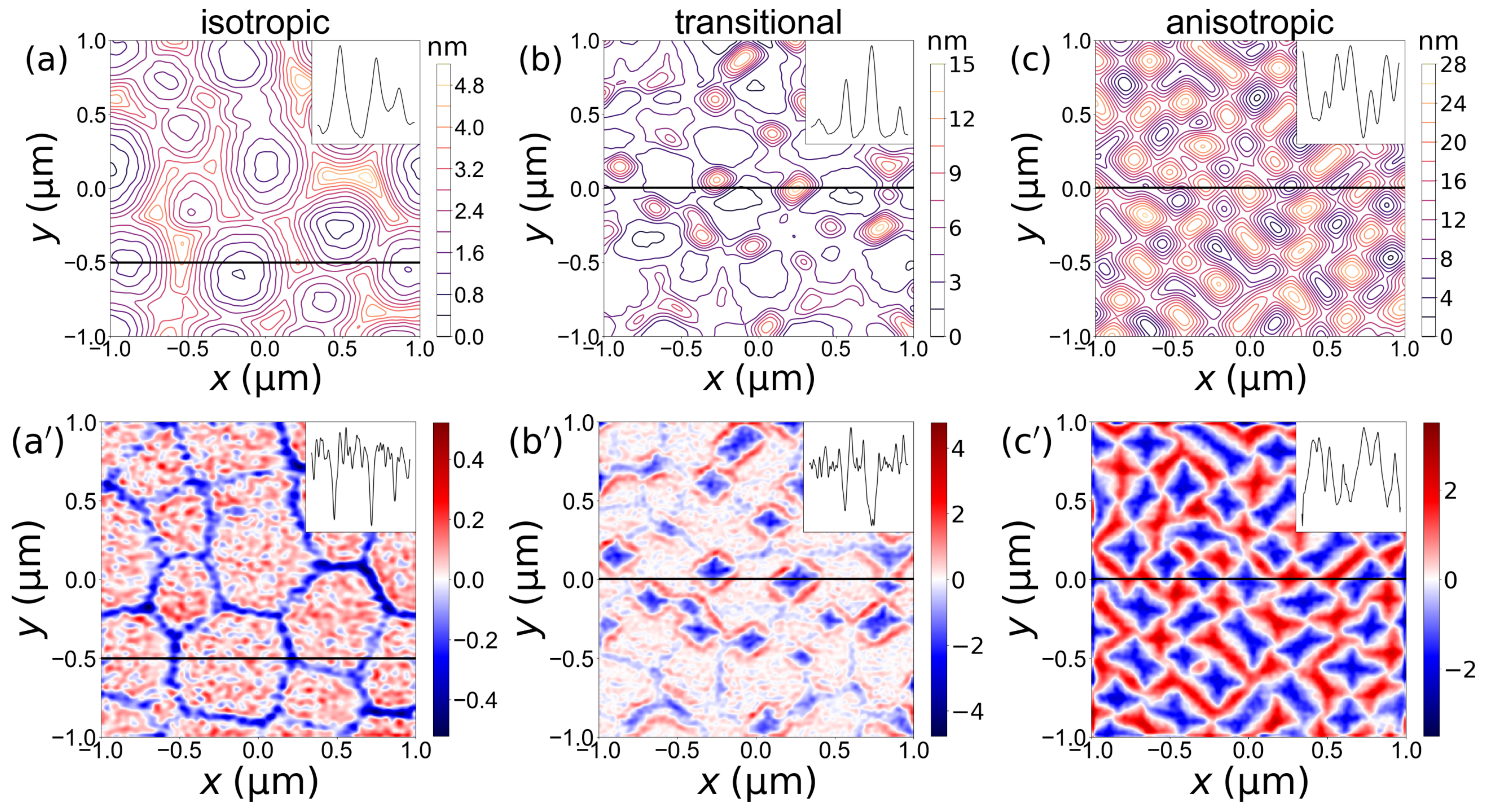}
\caption{Results obtained from analysis of the three surfaces shown in Fig.~\ref{fig:energy-series}. Panels (a)-(c) are contour plots of the surfaces, and (a')-(c') show the mean curvature of the surfaces. The units of the mean curvature are $1/\mathrm{\mu m}$. The insets are cross-sections along the black lines in the corresponding image.}
\label{fig:experiment_contourcurv}
\end{figure*}

\begin{figure*}
\includegraphics[width=\textwidth]{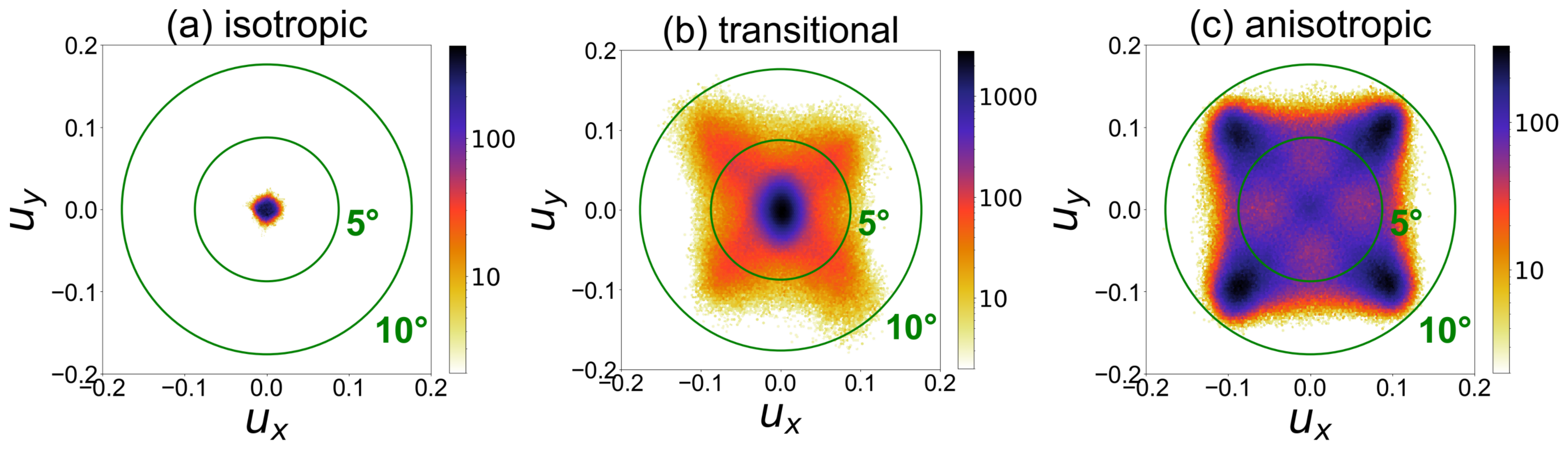}
\caption{Two-dimensional histograms of gradient of the surface height for the different types of pattern, as observed for irradiation at a sample temperature of \SI{380}{\degreeCelsius}, an ion flux of $1\times 10^{15}$ cm$^{-2}$s$^{-1}$, and an ion energy of (a) 400~eV, (b) 600~eV, (c) 800~eV. }
\label{fig:angle-distribution}
\end{figure*}

\subsection{Experimental Results
\label{ssct: Experimental Results}}

The AFM topographical images in Fig.~\ref{fig:temperature-series} show how the surface morphology resulting from irradiation with 1000~eV $\text{Ar}^{+}$ ions depends on the sample temperature. It has been demonstrated previously that the Ge(001) surface remains smooth for temperatures $T$ below the dynamic recrystallization temperature $T_R\cong$ \SI{250}{\degreeCelsius} and above \SI{500}{\degreeCelsius} \cite{Ou2013}. Within the temperature range $250~^{\circ}\text{C} < T < 500~^{\circ}\text{C}$, three different kinds of surface patterns are observed. The surface shows an anisotropic pattern of alternating rectangular pyramids and inverted rectangular pyramids with edges running along the $\langle 100 \rangle$ and $\langle 010 \rangle$ crystalline directions in the lower part of this range, as seen in Fig.~\ref{fig:temperature-series}(a). In the upper part of this temperature range, shallow basins with no obvious anisotropy form --- see Fig.~\ref{fig:temperature-series}(c). We refer to this as an isotropic pattern. Finally, at \SI{390}{\degreeCelsius}, we observe an intermediate pattern composed of small rectangular peaks roughly oriented along the $\left\langle100\right\rangle$ and $\left\langle010\right\rangle$ directions and large, shallow, isotropic basins, as in  Fig.~\ref{fig:temperature-series}(b). In this kind of pattern, structures characteristic of the anisotropic and the isotropic patterns coexist. The intermediate surface morphology was not observed by Ou {\it et al.}~\cite{Ou2013} because they did not carry out experiments at temperatures between $350~^{\circ}\text{C}$ and $430~^{\circ}\text{C}$.

AFM topographical images of Ge(001) surfaces irradiated at \SI{380}{\degreeCelsius} are shown in Fig.~\ref{fig:energy-series} for three different ion energies. At 400~eV, the pattern is isotropic, and is similar to the pattern observed at an ion energy of 1000~eV and a temperature of \SI{410}{\degreeCelsius}. An intermediate pattern was found for 400 eV, while an anisotropic pattern formed for ions with an energy of 800~eV.

\begin{figure*}
\includegraphics[width=\textwidth]{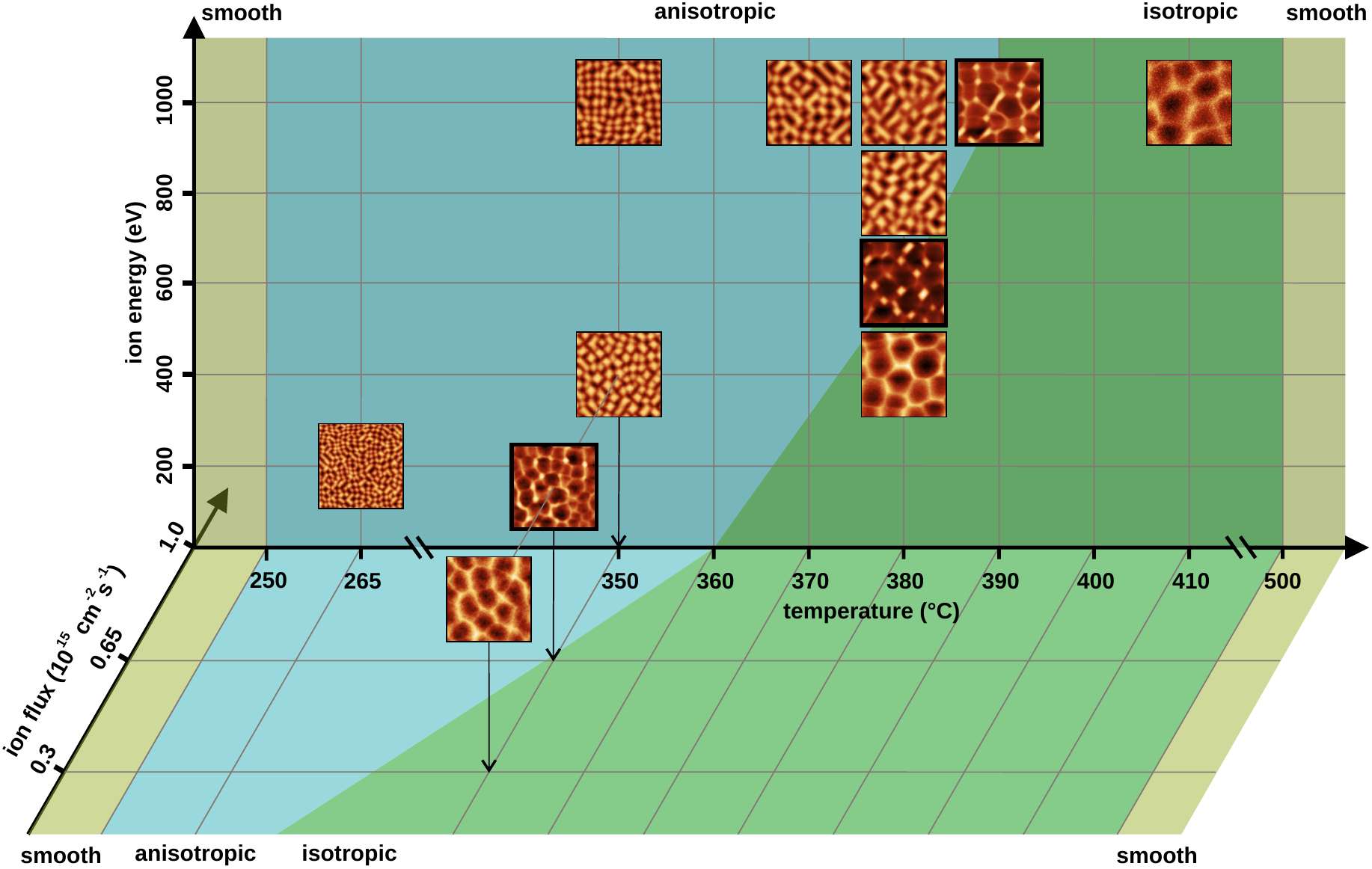}
\caption{Summary of the experimental results, showing AFM micrographs of the different observed morphologies at the respective positions in the parameter space of surface temperature, ion energy, and ion flux. Regimes for smoothing, anisotropic and isotropic patterning, respectively, are indicated by the differently colored areas.}
\label{fig:experiment_summary}
\end{figure*}

Figure~\ref{fig:flux-series} shows AFM topographical images of Ge(001) surfaces irradiated at a temperature of \SI{350}{\degreeCelsius} and an ion energy of 400~eV for three different ion fluxes:  $f = 3 \times 10^{14}\text{cm}^{-2}\text{s}^{-1}$, 
$6.5 \times 10^{14}\text{cm}^{-2}\text{s}^{-1}$
and $1 \times 10^{15}\text{cm}^{-2}\text{s}^{-1}$. As for increasing ion energy, we observe a transition from an isotropic to an intermediate and finally to an anisotropic pattern with increasing ion flux. Thus, sample temperature as well as ion energy and ion flux determine whether an isotropic, intermediate or anisotropic pattern is formed.

To probe the morphology of the isotropic, intermediate and anisotropic patterns in greater depth, we carried out additional analysis of the three surfaces shown in Fig.~\ref{fig:energy-series}. The top row of Figure \ref{fig:experiment_contourcurv} shows contour maps of the surfaces, and the bottom row shows the mean curvature $\nabla^2 u$ of the surfaces as a function of $x$ and $y$. The results in Fig.~\ref{fig:experiment_contourcurv} (a) and (a') indicate that the basins of the isotropic pattern are curved, but the maximum mean curvature is an order of magnitude smaller than in the intermediate and anisotropic cases.  
Figure \ref{fig:experiment_contourcurv} (b) is a contour map of
the intermediate surface and shows that isolated, sharp peaks with rectangular symmetry protrude from a landscape of shallow, rounded basins.   Figure~\ref{fig:experiment_contourcurv} (b'), on the other hand, shows that the mean curvature $\nabla^2 u$ is large in magnitude and negative at the apexes of the peaks and that the areas between the peaks are comparatively flat. Finally, Figs.~\ref{fig:experiment_contourcurv} (c) and (c') confirm that the anisotropic pattern consists of upright and inverted pyramids.

Figure~\ref{fig:angle-distribution} shows two-dimensional (2D) histograms of the gradient of the surface height for the three different types of patterns. The azimuthal positions of the four off-center maxima in the gradient histogram for the anisotropic pattern evidence the rectangular shape of the pyramidal structures, that are oriented along the $\langle100\rangle$ and $\langle010\rangle$ crystalline directions. The circular markers indicate the slopes corresponding to out-of-plane inclination angles of 5\degree and 10\degree, respectively. In the isotropic pattern all local slopes are close to zero, and therefore the 2D gradient histogram features only one central maximum at $0^{\circ}$. In contrast, the anisotropic pattern is dominated by the tilted side walls of pyramid-shaped nanostructures. Accordingly, the 2D gradient histogram features four maxima corresponding to the out-of-plane inclination angle of the side walls, reflecting the fourfold symmetry of the pattern morphology. The intermediate pattern combines the morphological characteristics of the isotropic and the anisotropic pattern. Its gradient histogram contains a central maximum as well as four off-center maxima, although they are less distinct and are at lower slopes than for the anisotropic pattern, indicating that the peak structures in this pattern have a similar four-fold symmetry albeit with less steep and less clearly faceted, but more rounded features.
The maxima in Fig.~\ref{fig:angle-distribution}(c) indicate an inclination of the pyramid walls of approximately $7.7^{\circ}$. A number of unrelated experimental studies of reverse epitaxial patterning on Ge(001) have reported different values for this inclination angle ranging from about $8^{\circ}$ to $11.5^{\circ}$ \cite{Ou2013, Erb2020, Erb2021}.

The experimental results are summarized in Fig.~\ref{fig:experiment_summary}, which shows AFM micrographs of the different observed morphologies at the respective positions in the parameter space of surface temperature, ion energy, and ion flux. We investigated the temperature range from 265 to \SI{410}{\degreeCelsius}, the energy range from 200 to 1000~eV, and the flux range from $0.3\times 10^{15}$ to $1\times 10^{15}\text{cm}^{-2}\text{s}^{-1}$. Earlier work showed that the surface smooths and consequently patterning does not occur for temperatures below \SI{250}{\degreeCelsius} or above \SI{500}{\degreeCelsius} \cite{Ou2013}. For all ion energies at temperatures within these limits, the isotropic pattern results from irradiation at higher temperatures than the anisotropic pattern. The temperature at which the transition from anisotropic to isotropic patterns occurs increases with increasing ion energy and with increasing flux. Roughly speaking, there are two different regimes: The first regime occurs for lower temperatures in combination with high ion fluxes and/or high ion energies and results in anisotropic patterns. The second regime, on the other hand, occurs for high temperatures together with low ion fluxes and/or low ion energies and results in isotropic patterns. Intermediate patterns are found between these two regimes.

\section{Theory}

\label{sec: theory}

\subsection{Equation of Motion}

\label{ssec: eom}

Consider the planar (001) surface of a germanium single crystal that is maintained at a temperature $T$ above the recrystallization temperature $T_R$.  We place the origin on the surface of the sample and orient the $z$ axis so that the sample occupies the region $z\le 0$.  In addition, we orient the $x$ axis so that it is parallel to the $\langle 110 \rangle$ direction.  

Suppose that the sample surface is perturbed slightly and then it is continuously bombarded with a normally incident, broad noble gas ion beam. We will employ a continuum description of the surface dynamics in which the position of an arbitrary point $\bm{r}$ on the solid surface is given by $\bm{r}=x\bm{\hat{x}}+y\bm{\hat{y}}+h(x,y,t)\bm{\hat{z}}$, where $h(x,y,t)$ is the height of the point above the $x-y$ plane at time $t$.

Let $-v_0$ be the velocity with which the surface would recede due to erosion by the ion beam if it were completely flat and set $h(x,y,t) = -v_0t + u(x,y,t)$, where $u$ is the deviation of the surface height from its unperturbed steady-state value.
The equation of motion (EOM) advanced by Ou~{\it et al.}~as a model of the time evolution of the Ge (001) surface is
\be
u_t = -A\nabla^2 u - B \nabla^2\nabla^2 u 
+ \nu(\partial_x u_x^3+\partial_y u_y^3) + r \nabla^2(\nabla u)^2,
\label{eom}
\ee 
where $A$, $B$, $\nu$ and $r$ are constants \cite{Ou2013}. The first term on the right-hand side of Eq.~(\ref{eom}) comes from the effect of the Ehrlich-Schwoebel (ES) barrier and curvature-dependent sputtering \cite{Sigmund1973,Bradley1988}. The origin of the second term is thermally activated diffusion.  This term has a stabilizing effect, i.e., it tends to smooth the surface. The third term describes the nonlinear contribution of the ES effect to the dynamics and it is fourfold rotationally invariant about the $z$ axis. The fourth and final term is known as the conserved Kuramoto-Sivashinsky (CKS) nonlinearity.  This term is frequently included in the EOM for an ion-bombarded surface because it leads to coarsening, and coarsening of the surface is often observed in experiments \cite{Kim2004,Castro2005,Munoz-Garcia2006,Munoz-Garcia2006b,Munoz-Garcia2008,Munoz-Garcia2014}.  In general, the CKS term includes contributions from both sputtering \cite{Kim2004} and mass redistribution \cite{Castro2005,Munoz-Garcia2006,Munoz-Garcia2006b,Munoz-Garcia2008,Munoz-Garcia2014}.
The parameters $A$, $B$ and $\nu$ are positive, but $r$ need only be real. 

Research carried out since the work of Ou~{\it et al.}~has shown that modifications of the EOM (\ref{eom}) are needed \cite{Erb2020,Erb2021}. In particular, experiments in which the Ge(001) surface was bombarded at oblique incidence established that the slope dependence of the sputter yield must be taken into account \cite{Erb2020}.  This leads to the appearance of an additional term proportional to $(\nabla u)^2$ on the right-hand side of Eq.~(\ref{eom}), the EOM for normal incidence bombardment.  This term is known as the Kuramoto-Sivashinsky (KS) nonlinearity.

It is well known that the sputter yield of a surface depends on its curvature \cite{Sigmund1973,Bradley1988}.  For normal-incidence ion bombardment of a sample, the leading order correction to the sputter yield that comes from the curvature dependence is proportional to the mean curvature $H\cong \frac12 \nabla^2 u$. This effect contributes to the first term on the right-hand side of Eq.~(\ref{eom}), as we have already noted. In addition, there is a higher order correction term to the sputter yield that is proportional to the squared mean curvature  $H^2\cong \frac14 (\nabla^2 u)^2$  \cite{Bradley2021b}. This term is not included in Eq.~(\ref{eom}). 

In many derivations of the EOM for an ion-bombarded surface, $u$ and $\nabla$ are in effect taken to be small and then an expansion is carried out in powers of these quantities. (For an explanation of why $\nabla$ can be treated as small, see Footnote \cite{exp}.)  Only terms up to a selected order in $u$ and $\nabla$ are retained in the expansion.  The CKS nonlinearity $r \nabla^2 (\nabla u)^2$ is second order in $u$ and fourth order in $\nabla$.  The term $(\nabla^2 u)^2$ is also second order in $u$ and fourth order in $\nabla$, and so for the sake of consistency, a term of this kind should be appended to the right-hand side of Eq.~(\ref{eom}).  The KS nonlinearity is second order in $u$ but is only second order in $\nabla$.  This means that it is in general larger than the CKS nonlinearity and so it must also be included in the EOM.  We therefore arrive at the generalized EOM
\bea
u_t  &=&-A\nabla^2 u - B \nabla^2\nabla^2 u +\lambda (\nabla u)^2 
 + r \nabla^2(\nabla u)^2 \nonumber\\
 &&+\alpha(\nabla^2 u)^2 + \nu\partial_x u_x^3 + \nu\partial_y u_y^3,
\label{gen eom}
\eea 
where $\alpha$ and $\lambda$ are the constant coefficients of the squared mean curvature (SMC) term and the KS nonlinearity, respectively.  

Additional generalizations of the EOM are possible.  For example, surface diffusion could be anisotropic.  This would lead to the appearance of a term proportional to $u_{xxyy}$ on the right-hand side of Eq.~(\ref{gen eom}). However, these generalizations are not needed to model the results of our experiments and so they will not be considered further here.

As we shall see, the SMC term $\alpha(\nabla^2 u)^2$ in Eq.~(\ref{gen eom}) is the key to the formation of the intermediate patterns --- it is what leads to the formation of isolated peaks on the solid surface.  The effect of this term is particularly evident if $r=\nu=0$. This special case of Eq.~(\ref{gen eom}) is known as the modified Kuramoto-Sivashinsky equation (MKSE).  Bernoff and Bertozzi proved rigorously that if $\alpha$ is nonzero, the MKSE produces spike singularities in finite time for a range of initial conditions of arbitrarily small amplitude \cite{Bernoff1995}.  These singularities are unphysical and must be controlled by a term that is not included in the MKSE.  As we will see shortly, the ES term in Eq.~(\ref{gen eom}) can prevent the formation of these singularities.  Roughly speaking, this is because it strongly inhibits the development of large surface slopes.

We can simplify our EOM (\ref{gen eom}) by 
introducing the dimensionless quantities $\tilde x\equiv (A/B)^{1/2} x$, $\tilde y\equiv (A/B)^{1/2} y$ and $\tilde t \equiv A^2 t/B$. Equation~(\ref{gen eom}) becomes
\bea
u_{\t t}  &=&-{\t\nabla}^2 u - {\t\nabla}^2{\t\nabla}^2 u +\t \lambda ({\t\nabla} u)^2 
 + \t r {\t\nabla}^2({\t\nabla} u)^2 \nonumber\\
 &&+\t \alpha({\t\nabla}^2 u)^2 + \t\nu\partial_{\t x} u_{\t x}^3 + \t\nu\partial_{\t y} u_{\t y}^3,
\label{semi eom}
\eea 
where $\t\lambda\equiv \lambda/A$, $\t r\equiv r/B$,
$\t\alpha\equiv \alpha/B$, and $\t\nu\equiv \nu/B$.  Because the values of the parameters $A$ and $B$ are unknown, we will carry out simulations of Eq.~(\ref{semi eom}) rather than of Eq.~(\ref{gen eom}).  Plots of $u$ versus the dimensionless lengths $\t x$ and $\t y$ will be given; if $A$ and $B$ were known; these plots could be rescaled to show $u$ as a function of $x$ and $y$.

\subsection{Simulation Method}

\label{ssec: method}

In our simulations of Eq.~(\ref{semi eom}), we used an $N\times N$ grid of points evenly spaced on the spatial domain with $-\t l\leq \t x \leq \t l$ and $-\t l\leq \t y \leq \t l$. Periodic boundary conditions were employed. All of the simulations were begun with a low amplitude spatial white noise initial condition. (The amplitude of the noise was chosen to be $10^{-3}$.) The numerical integrations were carried out using fourth-order Runge-Kutta exponential time differencing (ETDRK4) \cite{Cox2002,Kassam2005}. The linear terms were evaluated exactly in Fourier space, whereas the KS and SMC terms were approximated using finite differencing in real space. The CKS and ES terms were evaluated using both finite differencing in real space and a pseudo-spectral method. For example, the CKS term $\t r\t\nabla^2 (\t\nabla u)^2$ was approximated by calculating $(\t\nabla u)^2$ in real space using finite differencing, but the Laplacian of the result was then evaluated in Fourier space. Unless otherwise noted, the simulation parameters in the following results were $\t l=30\pi$ and $N=512$, and the time step was $\Delta\t t=0.01$. We checked numerical accuracy by verifying that increasing $N$ and decreasing $\Delta \t t$ did not affect the results significantly.

\subsection{Simulation Results}

\label{ssec: simulations}

\begin{figure*}
\includegraphics[width=\textwidth]{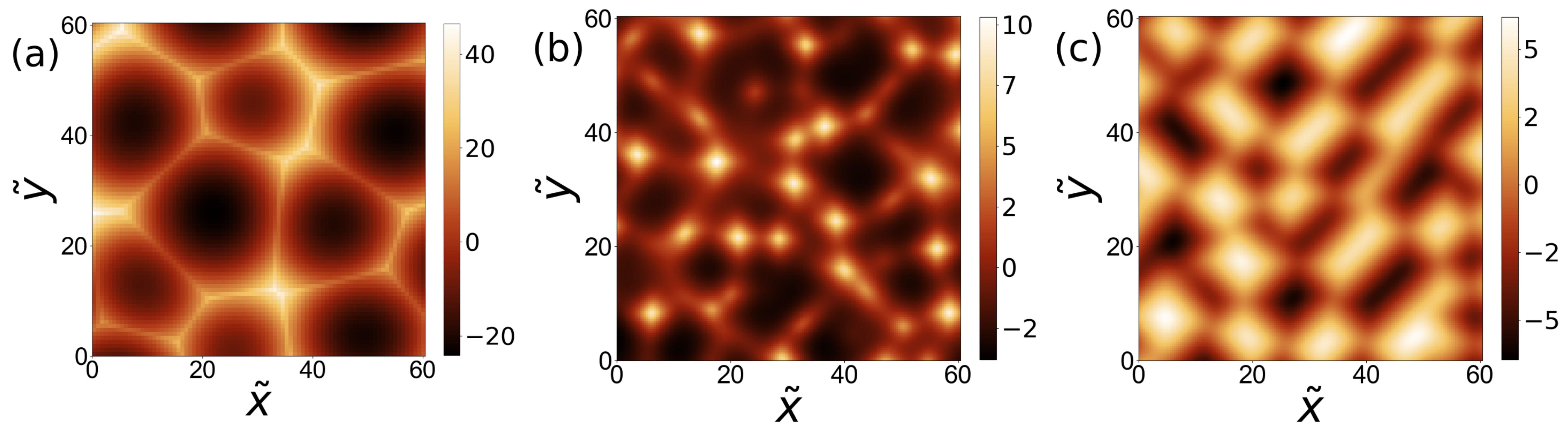}
\caption{The surface height $u$ at dimensionless time $\t t=200$ for three simulations of Eq.~(\ref{semi eom}) with different parameters. In (a), 
the parameter values were $\t r=1$ and $\t \lambda=\t \alpha=\t \nu=0$.
The result in (b) was obtained with $\t \lambda=-1$, $\t \alpha=\t \nu=1$ and $\t r=0$. In (c), we set 
$\t \nu=1$ and $\t \lambda=\t r=\t \alpha=0$.}
\label{fig:simulation_surfaces}
\end{figure*}

\begin{figure*}
\includegraphics[width=\textwidth]{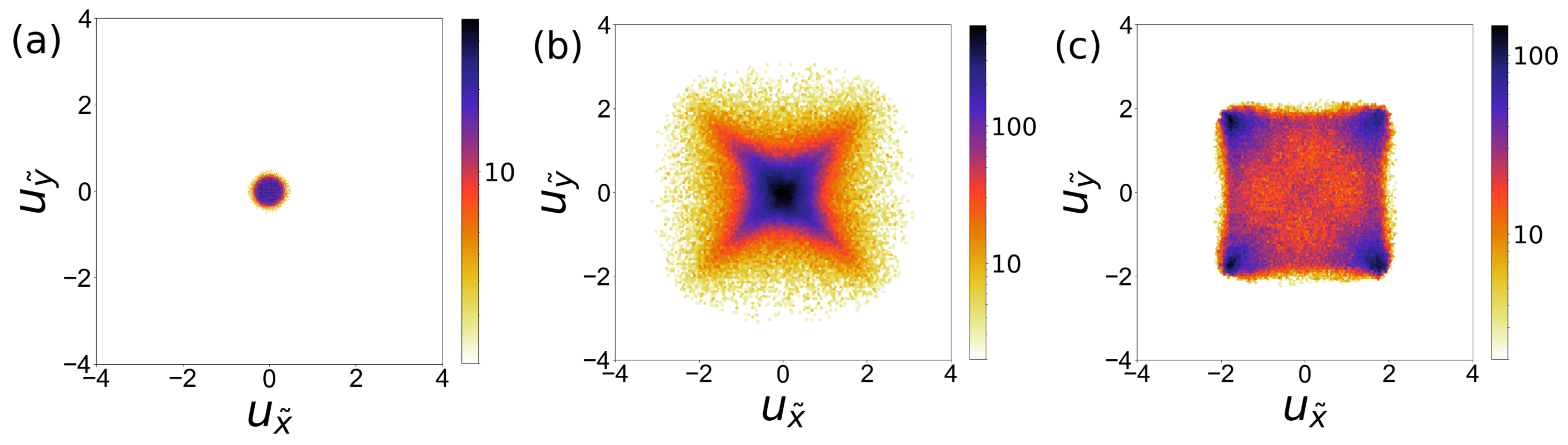}
\caption{Plots of the 2D gradient histogram at dimensionless time $\t t=100$ from three simulations of Eq.~(\ref{semi eom}) with different parameters. In (a), the parameter values were $\t r=10$ and $\t \lambda=\t \alpha=\t \nu=0$. The result in (b) was obtained for the parameter values $\t \lambda=-1$, $\t \alpha=\t \nu=1$ and $\t r=0$. In (c), $\t \nu=0.25$ and $\t \lambda=\t r=\t \alpha=0$.}
\label{fig:simulation_gradients}
\end{figure*}

The model advanced by Ou~{\it et al.}~is a special case of our EOM (\ref{semi eom}) with $\t\lambda=\t\alpha=0$.  Their model agrees well with their experimental observations for temperatures $T=260^\circ$C, $350^\circ$C and $430^\circ$C~\cite{Ou2013}.  Equation~(\ref{semi eom}) with $\t\lambda=\t\alpha=\t r=0$ produces patterns that are very similar to the anisotropic patterns observed by us at $T=370^\circ$C and earlier by Ou~{\it et al.}~at $T=260^\circ$C and $350^\circ$C: see Fig.~\ref{fig:simulation_surfaces} (c).  In this case, simulations produce surface patterns that consist of rectangular pyramids and inverted rectangular pyramids with their edges lying parallel to the $\langle 110 \rangle$ and $\langle 1\bar 10 \rangle$ directions when they are projected onto the $x-y$ plane. The pyramids also coarsen with time. The corresponding 2D gradient histogram distribution has four peaks arranged on the corners of a square [see Fig.~\ref{fig:simulation_gradients} (c)], just as in the experiments of Ou {\it et al.}~\cite{Ou2013} and in our own experiments (see Fig.~\ref{fig:angle-distribution}(c)).   

For $T=430^\circ$C, Ou~{\it et al.}~argued that the experimentally observed isotropic morphologies can be modeled well by Eq.~(\ref{semi eom}) with positive $\t r$ and with all other parameters set to zero. In our experiments, we obtained similar patterns for $T=410^\circ$C, as seen in Fig.~\ref{fig:temperature-series}(c).  The temperature $T$ in this case is high enough that the kink ES barrier has a negligible effect and consequently $\t\nu$ may be set to zero \cite{Ou2013}. 
Figures~\ref{fig:simulation_surfaces} (a) and \ref{fig:simulation_gradients} (a) show the result of a simulation for this case. A pattern of shallow basins separated by low ridges forms, and it coarsens in time. As for the experimental results in the isotropic patterning case, the 2D gradient histogram of the simulated surface exhibits only one central maximum, indicative of all slopes on this surface being close to zero. The EOM (\ref{semi eom}) is rotationally invariant when $\t\nu$ is zero, and so the use of the term isotropic for these patterns is fitting.

We now turn our attention to the intermediate patterns that form at $390^\circ$C, a temperature that lies between the lower temperatures where anisotropic patterns are found and the higher temperatures where isotropic patterns form.  The EOM advanced by Ou {\it et al.}~does not produce patterns of this kind.

As we have already noted, the presence of the isolated peaks in the intermediate patterns strongly suggests that the SMC term $\alpha(\nabla^2 u)^2$ is important in producing this kind of pattern formation. Because the experimentally observed peaks and 2D gradient histogram have fourfold rotational symmetry, a fourfold rotationally invariant ES term must play a prominent role as well.  This term is also needed to prevent the peaks --- which are incipient singularities --- from becoming true singularities. As noted earlier, the KS nonlinearity $\lambda(\nabla u)^2$ is of lower order in $u$ and $\bm{\nabla}$ than the SMC term, and so it likely must also be included in the EOM. 
We therefore simulated Eq.~(\ref{semi eom}) with nonzero values of $\t\alpha$, $\t\nu$ and $\t\lambda$ and with the one remaining parameter $\t r$ set to zero. As seen in Fig.~\ref{fig:simulation_surfaces} (b), for selected parameter values, the resulting surfaces look remarkably similar to the intermediate patterns we observed in our experiments, and, what is more, the corresponding 2D gradient histogram shown in Fig.~\ref{fig:simulation_gradients} (b) has nearly the same form as in the experiments (see Fig.~\ref{fig:angle-distribution})(b), i.e. it is characterized by a strong central maximum and less pronounced off-central maxima in a four-fold symmetric arrangement. These results confirm that the SMC term likely plays an important role in the formation of the intermediate patterns.

\begin{figure*}
\includegraphics[width=\textwidth]{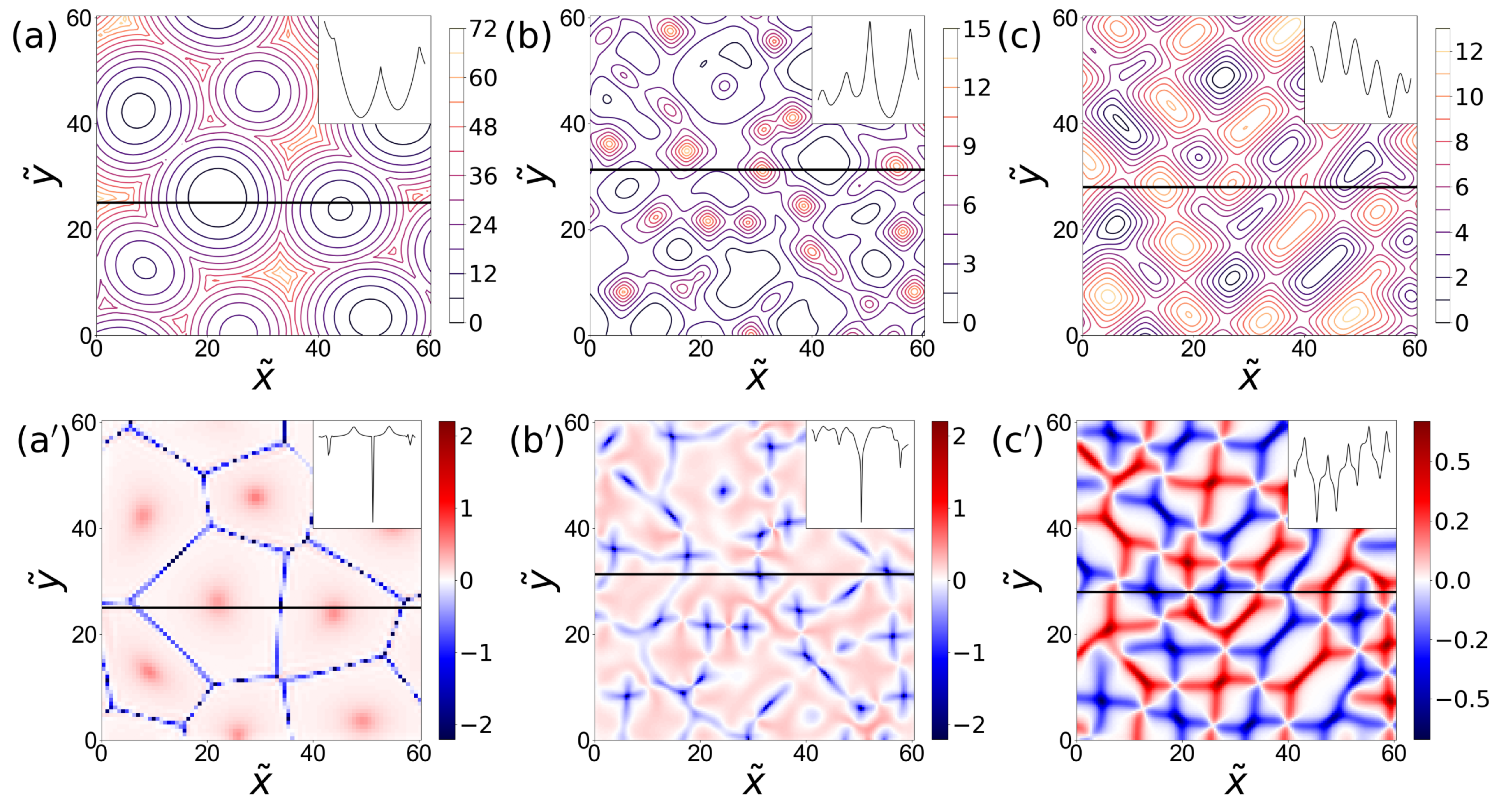}
\caption{Results obtained from analysis of the simulated surfaces shown in Fig.~\ref{fig:simulation_surfaces}. Panels (a)-(c) are contour plots of the surfaces, and (a')-(c') show the mean curvature of the surfaces. The insets are cross-sections along the black lines in the corresponding image.}
\label{fig:simulation_transitional}
\end{figure*}

We also produced contour plots and plots of the mean curvature for the three types of patterns we found in our simulations: see Fig.~\ref{fig:simulation_transitional}.  These plots are to be compared to the corresponding plots for the surface morphologies in our experiments, which are to be found in Fig.~\ref{fig:experiment_contourcurv}.  We see once again that the simulations reproduce the principal features of the experimentally observed patterns.  In particular, the contour plot for the intermediate pattern, which is shown in Fig.~\ref{fig:simulation_transitional} (b), confirms that widely separated, anisotropic peaks sit atop a landscape of shallow, rounded depressions, as in our experiments. There are, however, some differences.  
Comparing Fig.~\ref{fig:experiment_contourcurv} (c') with Fig.~\ref{fig:simulation_transitional} (c'), for example, we see that more of the surface has a small curvature in the theory than in the
experiments.  This means that in the theory, the facets are
flatter than in the experiments. This discrepancy could be remedied by including an additional term to the continuum equation that controls the degree of anisotropy of the ES term \cite{Erb2021}.  It was found that this term leads to better agreement between theory and experiment in an \emph{in situ} GISAXS study of ion-induced nanoscale pattern formation on crystalline Ge(001) \cite{Erb2021}.

The reader may reasonably ask why some terms are included in the EOM in certain temperature regimes but not others.  This question will be addressed in detail in Sec.~\ref{sec: discussion}.

\section{Discussion}
\label{sec: discussion}

In the simplest version of the viewpoint promoted by Levandovsky and Golubovi\'c \cite{Levandovsky2004} and Ou~{\it et al.}~\cite{Ou2013}, normal incidence ion bombardment of the (001) surface of a crystal is a kind of reverse epitaxy. The incident ions sputter away surface atoms and so produce vacancies in the crystal surface, and these vacancies diffuse on the crystal terraces until they attach to a step edge.  It is implicitly assumed that the rate vacancies are produced is the same at any point on the crystal surface. 
In the frame of reference that moves along with the solid surface, the continuum equation of motion conserves the volume of the solid.  The preceding statement also applies to the epitaxial growth of a solid, although, of course, during epitaxy, atoms are added to the solid rather than being removed. 

Contrary to this appealingly simple point of view, sputtering does not proceed at the same rate at all points on the solid surface except if the surface is planar.  Instead, the sputter yield depends on both the slope and curvature of the surface at an ion's point of impact \cite{Sigmund1973,Bradley1988}.  This means that the rate with which vacancies are produced on the surface depends on the surface morphology, and is generally different for different surface points. It also means that the dynamics need not conserve mass.

Ou~{\it et al.}~recognized that curvature dependent sputtering could affect the value of the coefficient of the term $-A\nabla^2 u$ in their EOM (\ref{eom}). However, both they and Levandovsky and Golubovi\'c did not include the slope dependence of the sputter yield, i.e., the KS nonlinearity $\lambda(\nabla u)^2$, in the EOM they advanced. Such a term does not appear in the EOM for the epitaxial growth of a solid under ideal growth conditions \cite{Lai1991},
but it does in general appear in the equation that governs the time evolution of a solid surface that is bombarded with a broad ion beam \cite{Cuerno1995}. The coefficient of the KS nonlinearity $\lambda$ is nonzero because the sputter yield depends on the local angle of ion incidence. When the KS nonlinearity is included in the EOM, the volume of the solid is not conserved in the co-moving frame of reference.  

Our results point to the crucial role that the KS term $\lambda(\nabla u)^2$ plays in the pattern formation that occurs when a Ge (001) surface with temperature $T>T_R$ is bombarded with a broad ion beam. As we saw in Sec.~\ref{sec: theory}, this term is needed if the model is to reproduce the intermediate patterns we observed.
This is in accord with the recent work of Erb {\it et al.}~\cite{Erb2020}, who found that their model could only produce patterns that resemble the ones they observed during oblique incidence bombardment of a crystalline Ge (001) surface if they incorporated an anisotropic KS nonlinearity into their model.

The SMC term also plays an essential role in the genesis of the intermediate patterns, since it is responsible for the formation of isolated peaks on the surface. This term would produce unphysical singularities if this were not prevented by the ES nonlinearity. The SMC term is a higher order correction term that stems from the curvature dependence of the sputter yield \cite{Bradley2021b}.  Our observation of the intermediate pattern appears to be the first experimental evidence for the effect of the SMC term in nanoscale pattern formation induced by ion bombardment.

As we have seen, our continuum model is able to produce patterns akin to the ones observed in our experiments for certain choices of the parameters.  Ideally, though, the parameters in the EOM (\ref{gen eom}) would be computed using the results of atomistic simulations for a given sample temperature, ion energy and fluence rather than being chosen in an {\it ad hoc} fashion to produce patterns like the ones we found in our experiments. This, however, is not possible at the present time.  If the target material is amorphous or a layer at the surface of the solid is amorphized by the ion bombardment, then the results of atomistic simulations can be input into the crater function formalism \cite{Harrison2014,Bradley2020,Hofsass2019} to yield estimates of the parameters in the continuum EOM.  However, a crater function formalism for crystalline target materials has not yet been developed.  This would be an interesting direction for future work, but it is beyond the scope of this paper.  

There is an additional complication that makes determining the parameters in our EOM (\ref{gen eom}) difficult: in addition to the ballistic effects caused by the incident ions, thermally activated effects play an important role in the behavior we observe in our experiments.  We have already noted that curvature-dependent sputtering and the ES barrier both affect the value of the coefficient $A$ in Eq.~(\ref{gen eom}).  Thermally activated surface self-diffusion clearly contributes to the coefficient $B$ of the term $-B\nabla^2\nabla^2 u$.  However, this term may not be as simple as one would think at first, since sputtering can also produce a term in the EOM that is proportional to $\nabla^2\nabla^2 u$ \cite{Makeev1997}. 
Similar considerations show that both ballistic and thermal effects influence the values of the parameters $r$ and $\nu$; only the coefficients $\lambda$ and $\alpha$ are independent of temperature because the corresponding terms in the EOM (\ref{gen eom}) come entirely from sputtering.  To further complicate matters, the relative importance of ballistic and thermal effects depends on the sample temperature, the ion energy and the ion flux.

Although we cannot compute the values of the coefficients in the EOM (\ref{gen eom}), we can make some plausible (if somewhat speculative) arguments about how they likely depend on the sample temperature $T$.
As $T$ increases, the thermal energy $k_\mathrm{B}T$ gets closer to the ES barrier, and the effective uphill mass current due to the ES barrier decreases as a consequence.  This means that $A$ and $\nu$ are both decreasing functions of $T$. Thermally activated surface diffusion has an Arrhenius temperature dependence, and so $B$ increases rapidly with $T$. Although the current understanding of the CKS term is incomplete, it too has a diffusive contribution and so its coefficient $r$ is also an increasing function of $T$ \cite{Villain1991,Raible2000,Haselwandter2007}.  Finally, as already noted, $\lambda$ and $\alpha$ do not depend on $T$. 
In the high temperature regime, $A$ is relatively small, and therefore the saturated amplitude of the pattern is comparatively small.  
This is why the surface width $w$ of the isotropic patterns is relatively small.  At the same time, $B$ and $r$ are relatively large. The characteristic lateral length scale of the patterns at early times, $(B/A)^{1/2}$, is therefore relatively large in this regime, as we observed in our experiments.  Moreover, because $r$ is large, as the amplitude of the pattern grows, the CKS term has a greater effect than the other terms in the EOM that are quadratic in $u$.  As a result, the KS and SMC terms have a negligible effect.  The effect of the ES nonlinearity, which is third order in $u$, is small because the surface width remains relatively small, and because its coefficient $\nu$ is small in the high temperature regime.  The upshot of this discussion is that Eq.~(\ref{gen eom}) with nonzero $A$, $B$ and $r$ and all the other parameters set to zero is expected to adequately model our experiments in the high temperature regime.  This conclusion is consistent with both our results and those of Ou {\it et al.}~\cite{Ou2013}. 

In the intermediate temperature regime, $r$ is smaller than in the high temperature regime.  Our results suggest that it may be set to zero in this temperature regime, at least as a first approximation. 
For appropriately chosen parameter values, the resulting EOM produces patterns that are quite similar to the intermediate patterns we found in our experiments, as we have seen.  Because $A$ is larger than in the low temperature regime, the surface width $w$ is larger.  The surface width is also increased by the SMC term, which produces isolated peaks on the surface. 

It remains to consider the low temperature regime.  $A$ and $\nu$ are largest in this temperature range.  
We introduce the dimensionless quantities $\bar x\equiv (A/B)^{1/2} x$, $\bar y\equiv (A/B)^{1/2} y$, $\bar t \equiv A^2 t/B$ and $\bar u = (\nu/B)^{1/2} u$. Equation~(\ref{gen eom}) becomes
\bea
{\bar u}_{\bar t}  &=&-{\bar\nabla}^2 {\bar u} - {\bar\nabla}^2{\bar\nabla}^2 {\bar u} +\bar \lambda ({\bar\nabla} {\bar u})^2 
 + \bar r {\bar\nabla}^2({\bar\nabla} {\bar u})^2 \nonumber\\
 &&+\bar \alpha({\bar\nabla}^2 {\bar u})^2 + \partial_{\bar x} {\bar u}_{\bar x}^3 + \partial_{\bar y} {\bar u}_{\bar y}^3,
\label{nodim eom}
\eea 
where $\bar\lambda\equiv (B/\nu)^{1/2}\lambda/A$,
$\bar\alpha\equiv (B\nu)^{-1/2}\alpha$ and $\bar r\equiv (B\nu)^{-1/2} r$.  When $A$ and $\nu$ are large, $\bar\lambda$,
$\bar\alpha$ and $\bar r$ are small and therefore  
the effect of the quadratic KS, SMC and CKS terms is negligible.  Thus, we conclude that Eq.~(\ref{gen eom}) with positive $A$, $B$ and $\nu$ and all of the other parameters set to zero provides a good model of the pattern formation in the low temperature regime, and this what both we and Ou {\it et al.}~found \cite{Ou2013}.

The arguments just given draw heavily upon our knowledge of how the various physical processes depend on the temperature.  At the present time, much less is known about the dependence of the parameters on the ion flux $f$ and energy $E$, and we cannot make comparable arguments about the dependence of the patterns on $f$ and $E$ as a result.  

\section{Summary and Conclusions
\label{sct:Conclusions}}

Three distinct surface morphologies can be observed on the crystalline Ge(001) surface after irradiation with a normally-incident argon ion beam; which morphology is observed depends on the ion energy and flux and on the surface temperature. The anisotropic morphology is close to being up-down symmetric and consists of alternating upright and inverted rectangular pyramids. 
The other two morphologies are up-down asymmetric.
The isotropic morphology is made up of shallow, rounded basins separated by low ridges.  In the intermediate morphology, on the other hand, isolated peaks with rectangular cross sections stand above a landscape of shallow, rounded basins.

All presented surface morphologies are found in the reverse epitaxy regime, where ion induced defects are annealed effectively due to high diffusivity and the surface remains crystalline. In fact, earlier works considering only the isotropic and anisotropic morphologies were able to convincingly reproduce both these pattern types with an EOM reduced to terms which describe exclusively diffusive processes \cite{Ou2013}. Consequently, the formation of these patterns was attributed to diffusion of vacancies and ad-atoms, with the effect of ion irradiation being merely the steady, spatially homogeneous production of these mobile species required to form the patterns.
Our present work shows that the effect of ion irradiation can not in general be reduced as far as previously thought, because an EOM with exclusively diffusive terms cannot reproduce the intermediate pattern. To account for our experimental results, it was necessary to extend the initial theory of Ou {\it et al.}~to include the effects of the slope and curvature dependence of the sputter yield. For a range of parameter values, the resulting model produces surface patterns that are remarkably similar to the intermediate morphologies we observed in our experiments. The characteristic isolated peaks in this type of pattern result from an erosive term in the equation of motion which relates to the squared mean curvature of the surface and would produce spike singularities if this were not averted by the ES effect.
This shows that slope- and curvature-dependent erosive aspects of ion irradiation can -- e.g. in case of the experimental conditions at the transition between the anisotropic and the isotropic patterning regime --  be influential enough to produce obvious deviations from purely diffusion-driven patterning. Their function is then more than the provision of mobile species, but they also contribute directly to the actual pattern morphology. Such erosive contributions should therefore be considered also in the reverse epitaxy patterning regime. In particular, it is reasonable to assume an influence of erosive effects on the patterning dynamics. This would be worth investigating in detail by studying how the observed types of pattern evolve with ion fluence.

\begin{acknowledgments}
Irradiations and topography imaging were carried out using instruments provided by the Ion Beam Center at the Helmholtz-Zentrum Dresden-Rossendorf e.V. -- a member of the Helmholtz Association.  R.M.B.'s research was supported by Grant DMR-2116753 awarded by the U.S.~National Science Foundation.
\end{acknowledgments}

\bibliography{Ge_transition-morphology.bib}

\begin{thebibliography}{41}
\bibitem [{exp()}]{exp}%
  \BibitemOpen
  \href@noop {} {}\bibinfo {note} {Suppose that $u(x,y)=U(X,Y)$, where $X\equiv
  \epsilon x$, $Y\equiv \epsilon y$ and $\e$ is small. This means that the
  surface height $u$ is a slowly varying function of $x$ and $y$. Let $\nabla_X
  \equiv \bm{\hat x}\partial_X + \bm{\hat y}\partial_Y$. We have $\nabla u =
  \e\nabla_X U$, and so $\nabla u$ is of order $\e$. Thus $\nabla = \e\nabla_X
  $ can be thought of as small.}\BibitemShut {Stop}%
\end{thebibliography}

\end{document}